\documentclass[10pt,conference]{IEEEtran}
\IEEEoverridecommandlockouts
\usepackage{comment}

\usepackage{cite}
\usepackage{amsmath,amssymb,amsfonts}
\usepackage{algorithmic}
\usepackage{graphicx}
\usepackage{textcomp}
\usepackage{xcolor}
\usepackage{booktabs}
\usepackage{listings}
\usepackage{longtable}
\usepackage{array}
\usepackage{xurl} %
\usepackage[hidelinks]{hyperref}

\usepackage{tikz}
\usetikzlibrary{positioning,arrows.meta,decorations.pathreplacing,calc}
\usepackage{soul}
\usepackage[most]{tcolorbox}

\graphicspath{{figures/}}

\newcommand{\authorcol}[1]{\makebox[0.3\textwidth][c]{#1}}

\definecolor{findingframe}{HTML}{808080}
\definecolor{findingfill}{HTML}{EFEFEF}

\newcommand{\goodnum}[1]{#1}
\newcommand{\findingbox}[2]{\begin{tcolorbox}[enhanced, colback=findingfill, colframe=findingframe, arc=2mm, boxrule=0.5pt, boxsep=0pt, left=4pt, right=4pt, top=3pt, bottom=3pt, before skip=5pt, after skip=5pt]\textbf{#1} #2\end{tcolorbox}}

\begin{document}

\title{From GPU to TPU with Coding Agents: What Works, What Fails, and What It Costs}

\author{\IEEEauthorblockN{\authorcol{Hui Sun}}
\IEEEauthorblockA{\textit{North Carolina State University}\\
Raleigh, NC, USA \\
hsun26@ncsu.edu}
\and
\IEEEauthorblockN{\authorcol{Anderson Uch\^oa}}
\IEEEauthorblockA{\textit{Federal University of Cear\'a}\\
Itapaj\'e, Cear\'a, Brazil \\
andersonuchoa@ufc.br}
\and
\IEEEauthorblockN{\authorcol{Wesley K. G. Assun\c{c}\~ao}}
\IEEEauthorblockA{\textit{North Carolina State University}\\
Raleigh, NC, USA \\
wguezas@ncsu.edu}
}

\maketitle

\begin{abstract}
A coding agent can migrate a machine learning workload, report success, and still change how the model trains. Rising compute costs, limited accelerator capacity, and vendor dependence are pushing organizations toward alternatives to graphics processing units (GPUs); a key question is whether agent-generated migrations to tensor processing units (TPUs) can be trusted and at what cost. We build a dataset of 148 PyTorch ML research projects and use Claude Code to migrate ten of them from NVIDIA GPUs to Google Cloud TPUs. We independently assess whether each migration runs, preserves numerical behavior, executes on the intended hardware, maintains the project's goal, and delivers acceptable performance. We also examine failures, human effort, and costs. Eight projects meet our project-goal criterion, and migrations require 2.4 human interventions on average. Of 70 execution attempts, 53 fail (76\%), and 26 of these failures (49\%) raise no error. The most frequent cause is repeated compilation and other execution inefficiencies, while interactions with third-party libraries can silently change the computation: in one four-chip migration, the agent reports success despite a change in gradient clipping that reduces accuracy from 86.53\% to 84.89\%. Execution costs vary widely: in one workload, TorchTPU, a PyTorch execution backend for TPUs, trains 1.86 times faster and 25\% cheaper per run than the GPU. For practitioners, these findings motivate independent reference runs, verification under the intended hardware configuration, and workload-specific acceptance and stopping criteria that account for both computational and human costs.
\end{abstract}


\section{Introduction}\label{sec:intro}

As AI applications become more widely deployed~\cite{maslej2025artificial}, organizations face growing concerns about computing costs~\cite{wen2025mindthegap}, energy efficiency~\cite{patel2026accelerators, minoza2025ecolyzer}, and dependence on hardware vendors whose capacity may be constrained~\cite{patel2026accelerators, hu2026jointtraining, kim2026hetccl}. Migrating machine learning (ML) workloads from GPUs to alternative accelerators such as TPUs offers greater deployment flexibility, but introduces engineering challenges involving software compatibility, behavioral preservation, and operational costs. The hard part of such a move is not the hardware but the engineering: a field study of deploying large models on a non-GPU accelerator concludes that \textit{``the total cost of ownership is dominated by engineering effort (patching, tuning, operating) rather than by acquiring the hardware''}~\cite{yu2026ascend}. 

Coding agents promise to cut this effort. Google used a multi-agent system to migrate production TensorFlow models to JAX, achieving a 6.4--8$\times$ speedup over manual migration~\cite{nikolov2026tf2jax}, and has released the TPU agent suite in early preview, a set of agents that helps automate onboarding, diagnostics, and inference performance tuning of PyTorch models on TPUs~\cite{google2026torchtpublog}; other work uses agents to deploy models onto edge NPUs~\cite{su2026aipc, su2026pytorchnpu}. Research tools, in contrast, mostly target single kernels and operators or framework translation, including work from Amazon~\cite{tang2026nkiagent, zhang2026accelopt}, Google~\cite{tschand2026jaxbench, liang2026maxcode, feng2026t2jbench}, and AMD~\cite{liu2026amdkernelvault, li2026npu}, and cross-vendor kernel generation~\cite{sereda2026kforge, wen2026multikernelbench}. However, reducing the effort required to adapt code does not establish whether an agent-generated migration can be trusted or whether its operational benefits justify the migration effort.

For engineering teams, adopting an agent-generated migration therefore raises a practical decision: \textit{what evidence is sufficient to accept the migrated workload, and at what cost?} Both are harder than they appear. In one of our migrations, the agent moved a fine-tuning project to a four-chip TPU; the program ran to completion without any error and the agent reported success, yet accuracy dropped from 86.53\% to 84.89\%, below even the baseline method of the original paper, because the project's pinned training library clipped gradients before averaging them across chips (Section~\ref{sec:res-hardware}). 

This failure illustrates a practical risk: a migration may pass execution checks while silently changing the behavior of an ML workload. Traditional repository-level migrations are commonly validated through compilation and existing test suites~\cite{ou2025rustrepotrans, liu2026migrationbench, ziftci2025migrating}. Such tests are often unavailable in ML research projects: among the 148 repositories we collected, only seven contained author-written tests. Even when a reference execution is available, comparing outputs across accelerators is challenging because legitimate numerical differences can arise from floating-point precision and operation order~\cite{feng2026t2jbench}. 

Moreover, agent-reported success is insufficient evidence: previous studies have documented agents claiming successful outcomes that independent assessments did not confirm~\cite{advani2026falsesuccess, salunkhe2026reclaim, lin2026reporescue}, including an accelerator deployment in which the target NPU remained inactive~\cite{su2026pytorchnpu}. Verification also has a cost: repeated training and evaluation runs consume accelerator resources, while stopping too early risks rejecting a viable migration. Organizations must therefore balance migration reliability, verification effort, and computational cost rather than treating successful execution as the sole acceptance criterion.

We investigate these challenges through an empirical study of coding-agent-driven GPU-to-TPU migration. We selected ten PyTorch research projects from a corpus of 148 repositories that modify or fine-tune foundation models. Of these, 10 have recorded migration attempts, and 9 produced completed TPU evaluation results. Research repositories provide a useful setting because their associated papers report quantitative outcomes against which migrated executions can be examined, although they do not capture all the operational requirements of production systems. Using Claude Code, we examine migrations from NVIDIA GPUs to Google Cloud TPUs, primarily through PyTorch/XLA, and additionally investigate TorchTPU in a case study.

We retain PyTorch while adapting the execution backend, supporting code, and dependencies when required. To evaluate migration reliability, we use five consistency dimensions (Section~\ref{sec:hierarchy}): executability (L0), numerical agreement (L1), execution integrity (L2), preservation of the project's goal (L3), and performance (L4). We distinguish verification outcomes from quantitative measurements and examine evidence from execution logs, numerical comparisons, runtime telemetry, and manual code inspection, rather than relying solely on the agent's assessment, since even state-of-the-art LLMs cannot reliably judge whether two programs are functionally equivalent~\cite{sun2026esem}.
We answer three research questions:
\begin{itemize}
\item \textbf{RQ1 (Consistency): Is the migrated project consistent with the original?} 
Eight of ten projects reached L3 under our human assessment, with an average of 2.4 human interventions per migration. However, agent-reported outcomes were not always supported by independent checks, showing that successful execution alone is insufficient evidence of correctness.
\item \textbf{RQ2 (Failure causes): Why do migrations fail, and which failures go unnoticed?} 
Overall, 49\% of failures raised no error. The most frequent cause involved lazy-tensor execution, including repeated recompilation and host synchronization. Library interactions could also silently alter computation, while the agent's diagnoses sometimes obscured the underlying cause.%
\item \textbf{RQ3 (Cost and failing fast): Is the migration worth it, and does the agent stop attempts that cannot succeed early?}
Agent token costs ranged from US\$2 to US\$23 per project, while a completed TPU run cost 0.47--3.6$\times$ its GPU counterpart, excluding failed attempts and verification. The agent stopped excessively slow executions but not silent behavioral drift, while fixed thresholds sometimes flagged acceptable differences as failures.
\end{itemize}

Two case studies show that hardware configuration and backend choice can change the outcome of the same migration (Sections~\ref{sec:res-hardware} and~\ref{sec:res-backend}). On four chips, the migrated LoRA+ silently lost 1.64\% accuracy while the agent reported success; with TorchTPU, its GPT-2 experiment trained 1.86 times faster and 25\% cheaper than on the GPU, with accuracy within 0.16\% of the GPU's.

This paper makes the following contributions to software engineering practice: \textit{(1) Migration reliability and failures:} empirical evidence from coding-agent-driven GPU-to-TPU migrations, using the L0--L4 assessment to identify discrepancies between agent-reported and independently inspected outcomes, silent failures, and their root causes; \textit{(2) Migration effort and cost:} reference data on token usage, migration attempts, human interventions, computational costs, and stopping behavior; \textit{(3) Actionable lessons:} recommendations for practitioners and tool builders on reference baselines, hardware- and dependency-aware verification, workload-specific acceptance criteria, and cost-aware stopping (Section~\ref{sec:implications}); and \textit{(4) Replication package:} screening data for 148 projects, migration instructions, and complete agent attempt trajectories~\cite{sun2026seipreplication}.

\section{Background and Related Work}\label{sec:background}

\textbf{PyTorch on TPUs.} In our study, PyTorch/XLA and TorchTPU provide two alternative execution backends for Cloud TPUs, with distinct execution models. PyTorch/XLA records training steps as lazy-tensor graphs compiled via XLA. This model introduces two distinct migration risks: unsupported operators may silently fall back to the CPU~\cite{wen2025mindthegap, pytorchxla2026migration}, while data-dependent tensor shapes can trigger repeated compilations that substantially increase execution time without raising an error. PyTorch/XLA exposes runtime metrics for CPU fallbacks and compilation events, which help detect unintended device execution and performance regressions but do not establish that training behavior has been preserved. In contrast, TorchTPU (\texttt{torch\_tpu}~0.2.0), the successor to PyTorch/XLA~\cite{basile2026torchtpu}, executes operations eagerly by default on a native \texttt{tpu} device while also supporting compilation. Neither backend guarantees bitwise agreement with GPU outputs. For engineering teams, backend selection therefore affects the verification required for a migration, even when the target hardware remains unchanged.

\textbf{Repository-Level Code Migration.}
Repository-level code migration is a long-standing research area, but it has mostly targeted migration between programming languages or versions, such as C-to-Rust translation~\cite{ou2025rustrepotrans, leanh2026rustprint, tadesse2026code}, multilingual repository translation~\cite{wang2025repotransbench}, and Java upgrades~\cite{liu2026migrationbench, ziftci2025migrating}. In these settings, migration success is commonly assessed through compilation and existing test suites~\cite{ou2025rustrepotrans, liu2026migrationbench, ziftci2025migrating}, although passing these checks does not guarantee behavioral equivalence. Our study instead examines migrations that retain PyTorch while changing the execution hardware and backend. Because the studied repositories rarely provide tests and numerical outputs need not match exactly across accelerators, acceptance requires evidence that relevant workload behavior is preserved. For research projects, the paper's key claim provides an additional workload-specific reference rather than a universal criterion for migration correctness.

\textbf{Agent-driven ML migration.} Beyond the kernel-level and framework-translation work discussed in Section~\ref{sec:intro}, agents also generate and optimize operators for Huawei Ascend NPUs~\cite{qiu2026agenticcann, wu2026ascendoptimizer}, scored kernel by kernel against hardware limits~\cite{gao2026cannbench}. Other studies, such as T2J-Bench~\cite{feng2026t2jbench}, already employ execution-grounded checks to evaluate repository-level transformations. Our focus is therefore on the challenges specific to agent-driven migration across hardware backends, including accelerator execution, training semantics, dependency interactions, and migration costs. Previous studies have also documented discrepancies between agent-reported success and independently verified outcomes~\cite{advani2026falsesuccess, salunkhe2026reclaim, lin2026reporescue}. We investigate how such discrepancies arise in GPU-to-TPU migrations, including failures that do not produce runtime errors.

\textbf{Feasibility awareness and early stopping.} Previous studies have investigated whether agents recognize infeasible tasks, often constructing them by removing prerequisites or tools or by mutating tests~\cite{luo2026agenticabstention, cheng2026feasigen, zhong2025impossiblebench}. In hardware migration, feasibility may become apparent only through resource-consuming attempts, while stopping rules can also prematurely reject viable migrations. We therefore examine early stopping as a practical trade-off between computational cost and the risk of abandoning a migration that could succeed.
\section{Methodology}\label{sec:design}

\subsection{Project Selection}\label{sec:selection}

We selected the studied projects in four steps. The data and scripts for every step, together with the prompts and reports of the search agents, are in the replication package~\cite{sun2026seipreplication}.

\textbf{Collection.} We defined four research areas in advance: parameter-efficient fine-tuning of language models (LMs); LM efficiency and compression; LM post-training (alignment, knowledge editing, unlearning, model merging, activation steering, continual learning, and instruction tuning); and vision foundation-model adaptation and LM architecture/inference modifications. Projects qualified if (IC1) their papers appeared at peer-reviewed venues in 2022--2026, (IC2) they had a public author-provided or official PyTorch repository, (IC3) they modified or fine-tuned a pretrained foundation model, and (IC4) at least one experiment fit on a single 16~GB accelerator. We excluded widely adopted projects (e.g., LoRA, DPO, GPTQ) to limit exposure to existing TPU ports and widely circulated solutions, without ruling out training-data overlap.
The search was assisted by LLM-based search agents, one per area, and every repository was checked against the GitHub API. This yielded 148 candidates.

\textbf{Screening.} A regex scan of each repository's Python code labels its primary migration challenge, taking the first match among custom CUDA extensions or kernels, linear algebra, dynamic shapes, distributed frameworks, and pure PyTorch. The label is a static prior used only to diversify the sample, not a measured outcome. We then scored each candidate with a feasibility penalty, which captures technical and resource constraints expected to complicate migration, such as experiments only at the 7B scale or outdated dependencies. %
We selected 100 projects with fixed quotas per area, balancing the primary challenges within each area and, among equally ranked projects, preferring lower penalties (details in the replication package~\cite{sun2026seipreplication}). This selection prioritizes coverage of different migration challenges and feasible evaluations rather than statistical representativeness.

\textbf{Sampling.} Of the 100 selected projects, 67 have no penalty. We drew ten, stratified by area and primary challenge, with a fixed random seed. The resulting sample supports comparisons across anticipated migration challenges but does not represent the distribution of ML workloads deployed in industry.

\textbf{Manual verification.} Two authors manually verified each of the ten sampled projects. They checked it against IC1--IC4, confirming the paper's venue in the official proceedings and that the repository is the authors' PyTorch implementation; checked that the repository does not already support TPUs, i.e., that the authors' own code contains no TPU or XLA execution path; and checked that its main experiment does not require gated models, registered datasets, or paid evaluation APIs. These checks ensure that the selected repositories require an actual migration and can be evaluated with the available infrastructure, while limiting the study to publicly accessible and resource-constrained workloads.

\subsection{The Consistency Assessment Framework}\label{sec:hierarchy}

We assess accelerator migrations along five complementary dimensions, distinguishing verification outcomes from quantitative measurements:
\textbf{L0 (Executable)} the project runs on the target accelerator; \textbf{L1 (Numerical)} how closely the computation reproduces the original's numerics, reported as the agreement on the final loss and on the project's own evaluation metric rather than as a verdict; \textbf{L2 (Execution)} the computation actually runs on the accelerator, without silent CPU fallback, repeated recompilation, or hidden errors; \textbf{L3 (Project Goal)} the original paper's key claim still holds under the accelerator's default numerics, or, when the original code already misses it on the GPU, the migrated result matches the GPU run; and \textbf{L4 (Performance)} the migration reaches L3, and its training run is faster and cheaper than the original code's on the GPU.

These dimensions form an ordered assessment rather than a strict five-level correctness hierarchy. L0 establishes that the experiment completes, while L2 verifies that it uses the intended accelerator and preserves relevant execution semantics. L3 assesses whether the project-specific goal remains supported, provided sufficient reference evidence is available. L1 is a quantitative measurement rather than an acceptance verdict, whereas L4 is a categorical judgment that we make after L3. In particular, repeated recompilation is treated as a performance problem rather than a correctness failure unless it prevents execution. A faster execution cannot compensate for a migration that changes the intended computation.

For reporting purposes, we distinguish the highest verified categorical level (L0, L2, L3, or L4) from the numerical agreement (L1) measurement. We also distinguish migration-induced failures from cases where the original GPU execution already fails to reproduce the published claim, and report inconclusive outcomes when the available evidence does not establish claim preservation. No universal numerical tolerance can reliably separate floating-point variation from defects across workloads~\cite{feng2026t2jbench}. Instead, we report the observed differences and leave acceptance thresholds to the requirements of the target application. The numerical thresholds used in our smoke-test protocol serve as triggers for further investigation, not definitive correctness criteria. For industrial adoption, this distinction allows teams to define workload-specific acceptance requirements while retaining explicit evidence of execution integrity, numerical behavior, and operational cost.

\subsection{Agent Setup}\label{sec:agent}

\begin{table*}[t]
\centering
\caption{Migration outcomes, consistency levels, effort, and costs across ten projects.}
\label{tab:results}
\scriptsize
\setlength{\tabcolsep}{1.45pt}
\begin{tabular}{@{}rllllrrrrrcccccrrrrr@{}}
\toprule
 & & & & & \multicolumn{3}{c}{Metric} & & & & \multicolumn{3}{c}{Level} & & Churn & \multicolumn{2}{c}{Agent} & Time (min) & Cost (\$) \\
\cmidrule(lr){6-8}\cmidrule(lr){12-14}\cmidrule(lr){17-18}
\# & Project & Challenge & Sub-experiment & Metric & Paper & GPU & TPU & Att. & FF & Human & 1st & Claim. & Meas. & Silent & (files, $+$/$-$) & \$ & h & GPU/TPU & GPU/TPU \\
\midrule
1 & MELoRA~\cite{ren2024melora} & CUDA & RoBERTa-b./MRPC & Acc. & 90.93 & 89.46 & 88.73 & 10 & 0 & 0 (0) & none & L3 & L4$^{\ddagger}$ & 3 & 3, $+$93/$-$1 & 6.89 & 0.97 & 35.9/12.0 & 0.51/0.24 \\
2 & UniPELT~\cite{mao2022unipelt} & Dyn. shape & BERT-b./STS-B & Spear. & 87.10 & 86.58 & 86.15 & 5 & 1 & 1 (0) & none & L3 & L3 & 2 & 11, $+$124/$-$15 & 5.44 & 1.04 & 13.6/8.5$^{*}$ & 0.19/0.17$^{*}$ \\
3 & ZO-Bench~\cite{zhang2024zobench} & CUDA & OPT-1.3B/SST-2 & Acc. & 92.20 & 92.78 & 92.09 & 10 & 1 & 3 (2) & none & L3 & L3 & 3 & 5, $+$196/$-$19 & 22.55 & 4.27 & 229.5/338.7 & 3.27/6.77 \\
4 & BiLLM~\cite{huang2024billm} & Lin. alg. & OPT-1.3B/WikiText2 & PPL & 69.97 & 77.08 & -- & 8 & 1 & 2 (0) & none & -- & none & 3 & 9, $+$116/$-$28 & 12.20 & 3.02 & 17.5/-- & 0.25/-- \\
5 & CoFi~\cite{xia2022cofi} & Dyn. shape & BERT-b./MRPC & Acc. & 82.60 & 82.60 & 79.17 & 8 & 2 & 3 (1) & none & L2 & L2 & 5 & 10, $+$298/$-$62 & 16.96 & 5.27 & 55.8/80.9 & 0.79/1.62 \\
6 & TDPO~\cite{zeng2024tdpo} & Dist. & Pythia-410M/HH$^{\S}$ & Acc. & 67.33 & 57.03 & 57.42 & 5 & 1 & 5 (4) & none & L3 & L4$^{\ddagger}$ & 5 & 5, $+$113/$-$23 & 13.97 & 2.72 & 658.2/268.9 & 9.36/5.38 \\
7 & LM-Steer~\cite{han2024lmsteer} & Lin. alg. & GPT-2/Sentiment & Pos. & 57.26 & 27.95 & 29.41 & 11 & 1 & 3 (1) & none & L3 & L3$^{\ddagger}$ & 3 & 12, $+$299/$-$18 & 15.65 & 4.24 & 271.3/146.4$^{\dagger}$ & 3.86/2.93$^{\dagger}$ \\
8 & MEND~\cite{mitchell2021mend} & Pure & BART-b./zsRE & ES & 89.0 & 89.22 & 89.24 & 3 & 1 & 1 (0) & L2 & L3 & L4 & 0 & 10, $+$139/$-$64 & 7.29 & 0.90 & 240.8/118.1 & 3.43/2.36 \\
9 & Tip-Adapter~\cite{zhang2022tipadapter} & Pure & CLIP RN50/Pets & Acc. & 89.70 & 89.42 & 89.67 & 5 & 1 & 3 (1) & none & L3 & L3 & 2 & 3, $+$46/$-$10 & 4.82 & 0.61 & 1.5/4.0 & 0.02/0.08 \\
10 & LoRA+~\cite{hayou2024loraplus} & CUDA & RoBERTa-b./MNLI-m & Acc. & 86.50 & 86.53 & 86.54 & 5 & 0 & 3 (1) & none & L3 & L3 & 0 & 6, $+$181/$-$1 & 2.00 & 0.48 & 78.3/88.6 & 1.11/1.77 \\
\bottomrule
\end{tabular}
\par\smallskip
\parbox{\textwidth}{\scriptsize Challenge: static prior (CUDA extension, linear algebra, dynamic shape, distributed, pure PyTorch). Acc./Spear./PPL/Pos./ES: accuracy, Spearman correlation, perplexity, positive share, editing success. FF: fail-fast stops. Human: interventions (operational ones in parentheses). L4: L3 reached, and training faster and cheaper than the original on the GPU, judged by us. --: no completed run. $^{*}$Early stopping after 43 (GPU) and 18 (TPU) epochs. $^{\dagger}$Includes generation and evaluation. $^{\ddagger}$The original code misses the paper's claim on the GPU; judged against that run. $^{\S}$Paper: Pythia-2.8B. Ratios in the text use unrounded times and costs.}
\end{table*}

Each migration uses a fresh Claude Code session with Claude Opus~5.5, the paper, repository, and replication-package instructions, following these steps:

\textbf{Preparation.} The agent creates a public GitHub repository that records the whole migration: its first commit is the original code at its upstream commit, and every later attempt and failure is recorded there, labeled with its level, its failure cause, and whether it was silent. The links to all migration repositories are in the replication package.

\textbf{Sub-experiment selection.} 
The agent selects one experiment, prioritizing: (1) relevance to the paper's key claim and availability of a reported result (L3 reference); (2) an executable script or configuration; (3) feasibility on one TPU v5e chip with HBM limitation and within approximately four hours; (4) accessible datasets and models; and (5) an available baseline for comparative claims. If the preferred experiment requires additional permissions, the agent requests approval rather than selecting a weaker alternative.

\textbf{Reference runs.} On an NVIDIA L4 GPU, the agent runs the unmodified code twice: a smoke test of at least 200 steps with a fixed seed and one evaluation, and a full run with the seed of the TPU full run, in parallel with the migration. The TPU smoke test uses matching settings and seed.

\textbf{Migration.} Attempt~0 runs the original code unchanged on the TPU. Each subsequent attempt is documented in a pull request recording the changes, expected consistency level, and observed outcome. Each detected failure, including one without an explicit error, is recorded in an issue with its cause, diagnosis, and proposed fix. Code changes are committed separately to preserve traceability.

\textbf{Level checks.}
After a full run, the agent records its claimed level before performing the checks: L0, successful completion and final metric; L1, numerical differences from the paper and GPU run; L2, absence of CPU fallback, repeated recompilation, NaNs, or changes to device placement, evaluation coverage, and optimization semantics; L3, preservation of the paper's key claim, also checked on the original GPU execution to distinguish pre-existing failures; and L4, throughput, cost per run, TPU utilization, and their ratios to the GPU. We independently inspect the recorded evidence rather than adopting the agent's verdict.

\textbf{Fail fast.} 
The agent stops if L0 is not reached within 10 attempts or three hours, or if it identifies an unavoidable blocker, such as an unsupported CUDA kernel. Before a full run, it compares TPU and GPU smoke tests under matching settings and waits for the experimenter's decision if the mean loss over the last 25\% of steps differs by more than 5\%, the evaluation metric by more than 1\% (one unit of the metric), or the loss trends disagree. Approval is also required if the estimated full run exceeds five hours. The agent records when it first suspected infeasibility.

\subsection{Data Collection and Manual Analysis}\label{sec:analysis}

For each migration, we collect GitHub pull requests (one per attempt, with the level the agent expects it to reach), failure issues (errors, cause labels, and agent diagnoses), agent logs and metrics, final code changes, and GPU/TPU execution logs, loss series, and evaluation curves.

We independently inspect the paper values extracted by the agent, migration results, and code changes across all attempts. The first author assigns each failure a root cause (Section~\ref{sec:rq2}) based on its error, the agent's diagnosis, and the subsequent fix. We include each label to supporting evidence in the replication package.

For \textbf{RQ1}, we assess the first migrated attempt and the final attempt using the L0--L4 checks and compare the published metrics with single full GPU and TPU runs under matching settings and the same seed. When the original GPU run reproduces the paper's key claim, we judge L3 against that run. When the original GPU run also misses the paper's claim, the failure is inherited from the original code; a TPU result that matches the GPU run then counts as L3 (marked $^{\ddagger}$ in Table~\ref{tab:results}), because the migration preserves the original behavior even though it cannot reproduce the published claim. We distinguish the agent's \emph{expected} level, stated before each attempt, from its final \emph{claimed} level, reported after execution but before its level checks. Falling short of an expectation indicates an optimistic prediction; \emph{false success} occurs only when the final claimed level exceeds our independently verified level.

For \textbf{RQ2}, we count failures by cause, separating silent failures. For \textbf{RQ3}, agent cost is token usage at list prices, while agent time spans from the initial instructions to the verified TPU full-run launch or fail-fast stop, excluding experimenter waiting and machine idle time. Per-run cost is training time (excluding setup) multiplied by the machine's list price. Code churn is the final diff against the original GPU version. We verify fail-fast stops against the evidence available at the time and count human interventions from agent conversations. 

\section{Results and Findings}\label{sec:results}

\subsection{Empirical Study}\label{sec:empirical}

\subsubsection{RQ1 Consistency}\label{sec:rq1}

Table~\ref{tab:results} summarizes the outcome of each migration. Under \emph{Metric}, \emph{Paper} is the published value, and \emph{GPU} and \emph{TPU} are single full runs of the original and the migrated code with the same seed; the GPU--TPU difference shows what the migration changes, within run-to-run variation, and the Paper--GPU difference shows whether the original code reproduces the paper. Under \emph{Level}, \emph{1st} is the level of attempt~1, the first run of the agent's modified code, judged on its smoke test; \emph{Claim.} is the level the agent claimed after its final full run, and \emph{Meas.} the level we determined. A claim above the measured level is a false success; L4 is judged by us after the fact, so the agent claims at most L3. \emph{Att.} counts all attempts, including attempt~0, which runs the original code unchanged; \emph{Silent} counts failed attempts that raised no error; and \emph{Time} and \emph{Cost} give the training time and the cost of one full run on the GPU and the TPU.

\textbf{Final level.} Of the 10 projects, 8 reach L3, 3 (MELoRA, TDPO, and MEND) also reaching L4; 1 stops at L2; and 1 fails to reach L0 within because our 10 times attempt constrains. For MELoRA, LM-Steer, and TDPO, L3 reflects preservation of the original GPU behavior rather than reproduction of the published claims: GPU and TPU results closely match, but both fall below the paper's reference ($^{\ddagger}$ in Table~\ref{tab:results}). CoFi reaches only L2 because its TPU run selects a smaller pruned model and falls below the baseline
For practitioners, a migration should be accepted against a verified GPU reference and the application's quality requirements; matching the original run alone is not enough.

\textbf{First and final attempt.} Only MEND reached L2 on its first migrated attempt, after fixing a hard-coded CUDA device identified by attempt~0; all other projects failed to reach L0. Across ten projects, we observed a median of 6.5 attempts (range 3--11) before completion or stopping. For practitioners, these results highlight the need to budget for iterative debugging and validation rather than expecting a one-shot migration.

\textbf{Numerical agreement (L1).} Across the 9 projects reaching L0, GPU--TPU metrics differ by a median of 0.43\% (range 0.01\%--3.43\%), with differences below 1\% in 7 projects; the median TPU-to-paper difference is 0.95\%. As GPU-to-paper gaps often exceed GPU--TPU differences, practitioners should establish a GPU reference before attributing discrepancies to migration and assess the results against workload-specific acceptance criteria.

\textbf{Performance (L4).} The full-run time after migration is a median of 0.62 times that before (range 0.33 to 2.58); five projects run faster after migration and four slower. Only MELoRA, TDPO, and MEND reach L4: LM-Steer's run is shorter only because its generation and evaluation are faster, and UniPELT's TPU run is shorter only because early stopping ended it after 18 rather than 43 epochs. Performance varies across workloads and backends, even without agent-driven tuning (Sections~\ref{sec:res-hardware} and~\ref{sec:res-backend}). For practitioners, these results show that successful migration does not guarantee faster execution: performance gains must be verified under comparable training conditions before making deployment decisions.

\textbf{Expected level and false success.} The agent frequently overestimated its pre-attempt outcomes: 40 of 57 attempts (70\%) fell short of their expected levels, or 36 (63\%) when the final runs of MELoRA, TDPO, and LM-Steer count as L3 (Table~\ref{tab:results}). BiLLM illustrates this optimism: the agent repeatedly expected L0 but never reached it, instead requesting human guidance before exhausting the budget. However, after full execution, the agent's final claims never exceeded our assessments in the 9 completed projects: they matched in 6 and were lower in 3, because the agent claims at most L3. CoFi, for example, expected L3 before its full run but correctly claimed L2 afterwards. Nevertheless, the four-chip case study revealed a false success: the agent reported a successful migration despite a silent accuracy regression (Section~\ref{sec:res-hardware}). For practitioners, these findings distinguish unreliable intermediate predictions from more accurate final assessments, while demonstrating that even final success claims require independent verification under the intended hardware configuration.

\findingbox{Finding (RQ1).}{Eight of ten projects reach L3, including three that preserve GPU behavior without reproducing their papers' claims. Migrations require 2.4 human interventions on average, highlighting the need for human oversight and independent, workload-specific verification before acceptance.}

\subsubsection{RQ2: Failure Causes}\label{sec:rq2}

\textbf{Classification.} To understand why migrations fail and which problems escape ordinary execution checks, we analyze both the underlying causes and whether failures raise an error. We classify failed attempts into seven non-exclusive root-cause categories (Table~\ref{tab:causes}), based on errors, agent diagnoses, and corrective code changes, while distinguishing pre-existing problems and false alarms. Across ten projects, 53 of 70 attempts failed (76\%), including 26 (49\%) without explicit errors. The agent predominantly labeled failures as ``device'' (24), ``numerics'' (8), or ``other'' (7), without explicitly distinguishing lazy-tensor execution or problems inherited from the original code. The replication package provides the complete failure classifications and supporting evidence.

\begin{table}[t]
\centering
\caption{Root causes and diagnostic symptoms of failed attempts (non-exclusive counts).}
\label{tab:causes}
\scriptsize
\setlength{\tabcolsep}{2pt}
\begin{tabular}{@{}>{\raggedright\arraybackslash}p{2.0cm}>{\raggedright\arraybackslash}p{4.75cm}r@{}}
\toprule
\textbf{Category} & \textbf{Cause and diagnostic symptoms} & \textbf{Count} \\
\midrule
C1 Environment/dependencies &
Incompatible pinned libraries or changed data/model locations; import, API, and package errors. &
7 \\

C2 CUDA/device assumptions &
GPU-specific calls (\texttt{nvidia-smi}, \texttt{--cuda}), CPU-resident tensors, or optimizer created before device transfer. &
8 \\

C3 Library interactions &
HF Trainer, PEFT, or accelerate alter parameter or checkpoint handling; stale parameters, incorrect checkpoint formats, or restoration errors. &
5 \\

C4 Lazy-tensor execution &
Dynamic shapes, host synchronization, or graph growth; repeated recompilation, slow steps, or increasing host memory without errors. &
13 \\

C5 Operators/numerics &
Unsupported operators, CPU fallbacks, rejected mixed precision, or numerical differences (e.g., sigmoid tail). &
7 \\

C6 Memory capacity &
Workload exceeds 16~GB HBM; out-of-memory errors. &
3 \\

C7 Infrastructure/harness &
External failures, including disk exhaustion or undetected GPU/TPU devices. &
4 \\
\midrule
Not attributed to migration &
Original code misses the paper's claim, smoke-test setup introduces differences, or failure attribution is inconclusive. &
7 \\
\bottomrule
\end{tabular}
\end{table}

\textbf{Environment and device assumptions (C1--C2).} Environment drift (7) and hard-coded CUDA assumptions (8) frequently arise in early migration attempts. Unlike silent failures, they typically produce errors that help identify incompatible dependencies or incorrect device handling. Although these problems require code changes, their explicit symptoms make them more straightforward to diagnose.

\textbf{Lazy-tensor execution (C4).} Once initial compatibility problems are resolved, successful execution may still conceal substantial performance degradation. Lazy-tensor execution is the most frequent cause, accounting for 13 failures across seven projects, all without errors. Repeated recompilation and host synchronization can make training steps tens of times slower or increase host memory consumption while the program continues running.

\textbf{Third-party library interactions (C3).} A different and potentially more consequential failure mode arises when library interactions silently change the computation itself. Among five C3 failures, CoFi retained stale parameter copies, while MELoRA failed to reload its best checkpoint. Both completed without errors, and their effects became apparent only through comparison with the GPU reference. Other C3 failures produced explicit errors, including replaced parameter objects in LoRA+ and checkpoint restoration problems in Tip-Adapter. These cases illustrate that adapting application code is insufficient when unchanged dependencies can alter training or evaluation behavior.

\textbf{Agent diagnoses and static predictions.} These failure mechanisms also expose limitations in automated diagnosis. The agent's broad labels obscure recompilation, host synchronization, and pre-existing reproduction problems, potentially directing debugging toward symptoms rather than underlying causes. Similarly, the static challenge identified during project selection matches the most frequent observed cause, including ties, in only three of eight projects with a predicted challenge.

\findingbox{Finding (RQ2).}{Nearly half of failed attempts (49\%) raise no error. Lazy-tensor execution is the most frequent cause (13 failures across seven projects), while third-party interactions can silently alter computation. For engineering teams, reliable migration acceptance requires runtime monitoring and behavioral verification beyond successful execution and agent-generated diagnoses.}

\subsubsection{RQ3: Cost and Failing Fast}\label{sec:rq3}

\begin{figure*}[t]
\centering
\includegraphics[width=0.8\textwidth]{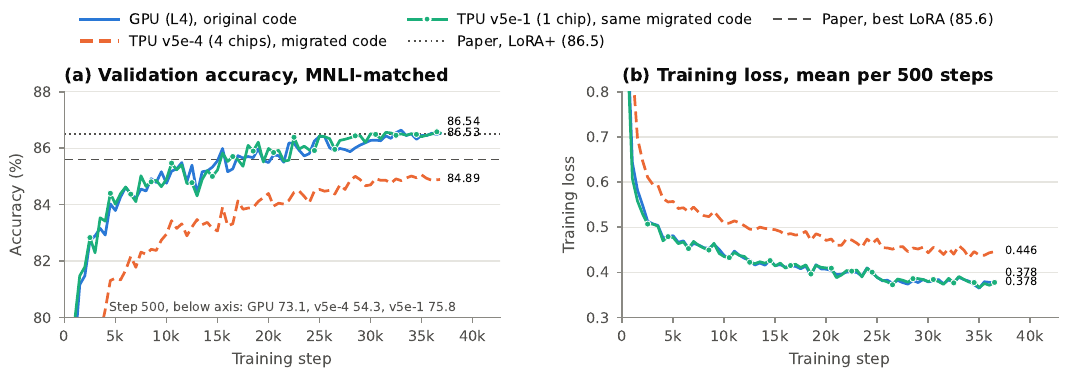}
\caption{LoRA+ on RoBERTa-base and MNLI with identical hyperparameters (global batch size 32, 36,816 steps): the original code on an NVIDIA L4 GPU, and the agent's migrated code on a four-chip TPU v5e-4 (eight examples per chip) and on a single-chip TPU v5e-1. (a) Validation accuracy every 500 steps, with the paper's LoRA+ and best tuned LoRA results as references. (b) Mean training loss over each 500-step interval.}
\label{fig:case-hardware}
\end{figure*}

\textbf{Why failing fast matters.} ML migrations incur costs beyond code generation: training and evaluation consume accelerator resources, and unsuccessful attempts may require repeated execution. In industry, these costs can accumulate as models are retrained frequently~\cite{he2014practical}.In our study, smoke tests take minutes to tens of minutes, while full TPU runs can take hours (Table~\ref{tab:results}). Identifying an unsuccessful attempt before a full run can therefore reduce wasted computation, but stopping prematurely may discard a viable migration. Failing fast is thus a trade-off between verification cost and the risk of incorrect rejection.

\textbf{Cost.} Across the ten projects, agent token charges have a median of \$9.74 (range \$2.00--\$22.55), with a median migration time of 1.88 hours. Median code churn is 7.5 files and $+$131.5/$-$18.5 lines. However, token charges represent only part of migration cost. Among the nine completed TPU runs, five cost less than their GPU counterparts and four cost more (Table~\ref{tab:results}). These comparisons also depend on equivalent workloads: UniPELT's lower TPU cost partly reflects earlier stopping. Moreover, failed attempts, validation runs, and human interventions add costs beyond a single successful execution. Consequently, low agent token costs do not establish that a migration is economically advantageous.

\textbf{Failing fast.} Across ten projects, the three stopping rules (Section~\ref{sec:agent}) triggered 9 distinct stops: 1 budget stop, 7 smoke-test stops, and 3 time-estimate stops, two of which triggered both of the latter rules. Only one of the seven smoke-test stops detected a migration-induced error; the other six were false alarms. The budget stop could not be assessed, and time-estimate stops flagged cost rather than incorrect computation. The agent also ended 15 attempts on its own. The following cases illustrate why stopping decisions require more than fixed thresholds.

\textbf{Examples.} In \emph{CoFi}~\cite{xia2022cofi}, the smoke-test gate triggered twice. The first stop correctly flagged pruning and fine-tuning losses 13\% and 24\% above the GPU reference, traced to stale CPU parameter copies and a 66\% discrepancy in the tail of \texttt{torch.sigmoid}. The second stop was a false alarm: a 5.3\% loss gap resulted from independently trained teacher models and fell to 0.29\% when both executions used the same teacher. Thus, the reference setup itself can trigger a rejection.

In \emph{UniPELT}~\cite{mao2022unipelt}, the gate stopped a migration because TPU training loss was 7.15\% \emph{lower} than GPU loss, despite an evaluation difference of only $-$0.28\%. After human approval, the full run reached L3. This false alarm illustrates the limitation of treating improvements and regressions symmetrically.

In \emph{LM-Steer}~\cite{han2024lmsteer}, a 2.8\% difference in sampled positive generations triggered the numerical gate, while the estimated full run exceeded five hours. A deterministic comparison showed matching next-token distributions, indicating that sampling variability caused the numerical alert. The time warning, however, prompted optimizations that reduced the estimated runtime from 14.4 to four hours. Here, the numerical gate was misleading, whereas the time estimate supported a useful intervention.

\textbf{Agent-initiated stops.} Beyond the prescribed rules, the agent independently stopped attempts with excessive projected runtime. In UniPELT, evaluation took 6.5~s per batch, suggesting a 12-hour run; in LM-Steer, generation remained on the first prompt for over ten minutes. Neither case raised an error. These decisions show that the agent can recognize impractical execution times, but we observed no comparable autonomous stopping in response to silent numerical or behavioral drift. Detecting such failures depended on the prescribed smoke-test checks and subsequent verification.

\findingbox{Finding (RQ3).}{Agent token costs are modest, but migration expenses also include accelerator runs and human intervention. The agent autonomously stops excessively slow attempts, while numerical gates detect additional failures but produce six false alarms among seven stops. Cost-effective migration therefore requires workload-aware stopping criteria and human review of ambiguous outcomes, rather than fixed thresholds alone.}

\subsection{Case Study: Single- vs Multi-chip TPU Migration}\label{sec:res-hardware}

\begin{figure*}[t]
\centering
\includegraphics[width=0.8\textwidth]{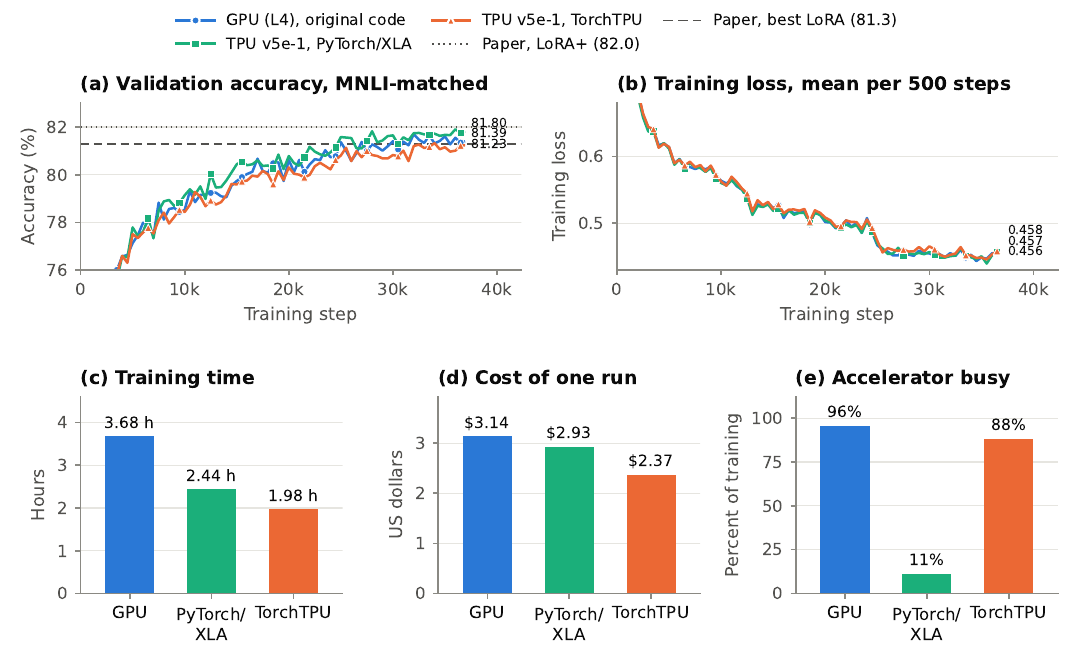}
\caption{
GPT-2 LoRA+ on MNLI (global batch size 32, 36,816 steps): original NVIDIA L4 GPU execution versus PyTorch/XLA and TorchTPU on a single-chip TPU v5e-1. (a) Validation accuracy every 500 steps, with published LoRA+ and tuned LoRA references (earlier evaluations fall below the axis); (b) mean training loss per 500 steps; (c) training time including evaluations; (d) per-run cost at on-demand prices (L4 \$0.8536/h; v5e-1 \$1.20/h); and (e) GPU utilization (\texttt{nvidia-smi}) and TPU duty cycle (\texttt{tpu-info}), which are not directly equivalent metrics.}
\label{fig:pilot-targets}
\end{figure*}

\textbf{Target and design.} This case study uses LoRA+~\cite{hayou2024loraplus}, whose key claim is that, once both methods are tuned, LoRA+ (86.5\%) outperforms LoRA (85.6\%) by \goodnum{$0.9$ percentage points}. Our GPU run of the original code reaches 86.53\% and serves as the reference: for L3, the TPU result must exceed it. We keep every hyperparameter (global batch 32, 36,816 steps) and vary only the chip count: on a single-chip v5e-1, all 32 examples of a step are on one chip; on a four-chip v5e-4, the agent's launcher runs one process per chip with eight examples each.

\textbf{Results.} Despite identical training settings, the two TPU configurations produce substantially different outcomes (Fig.~\ref{fig:case-hardware}). On v5e-1, the migrated code closely tracks the GPU, ending at 86.54\% (GPU: 86.53\%), remaining within 0.62\% of the GPU at every evaluation from step 2,000 onward, and reaching the same final training loss (0.378). On v5e-4, accuracy is already lower at step 500 (54.3\% versus 73.1\%), remains below the GPU at all 73 evaluation points, and ends at 84.89\%, with higher training loss (0.446). The 1.64-percentage-point drop exceeds the paper's reported 0.9-point advantage and places the migrated result below the published LoRA baseline (85.6\%). Thus, the four-chip execution fails our L3 assessment despite completing successfully.

\textbf{Migration on one chip v5e-1.} After resolving four blockers, including a torch\_xla upgrade and a checkpoint problem in the pinned Hugging Face Trainer (five files, $+$137/$-$1 lines), the migration reaches L3 on one chip. It is not attractive for L4: throughput is 0.89 times the GPU's and one run costs 1.59 times as much, with the TPU busy only about 13\% of a sampled interval (GPU: 91\%).

\textbf{Silent failure on four chips v5e-4.} Unlike the main migration protocol, this case study did not require the agent to compare TPU and GPU loss or evaluation metrics before the full run. On v5e-4, the program completed with exit code 0 and no error or warning, and the agent reported a successful migration. Yet manual inspection revealed that the distributed execution changed the optimization procedure. The pinned \texttt{accelerate}~0.27.2 has no XLA-specific gradient-clipping path, so each chip clips the gradient of its own 8 examples to norm 1.0 \emph{before} the cross-chip average, whereas the GPU clips the gradient of all 32 examples once. Clipping smaller per-chip gradients before averaging changes the resulting update and provides a plausible explanation for the slower convergence observed in Fig.~\ref{fig:case-hardware}. The agent's split preserved the global batch but did not check the order of clipping and cross-chip averaging, so on four chips the migrated code no longer computes the original optimization. This illustrates a critical migration risk: preserving configuration values and completing training does not ensure equivalent distributed training semantics. Asked to explain the gap, the agent blamed the TPU's default bfloat16 matrix multiplications, but the single-chip run uses the same numerics and matches the GPU, which rules this explanation out.

\findingbox{Finding.}{A migration that preserves results on one TPU chip can silently change optimization when distributed across multiple chips. Here, per-chip gradient clipping altered training behavior, while the agent reported success and misdiagnosed the regression. Reliable migration acceptance therefore requires verification under the intended hardware configuration, including distributed training semantics and dependency behavior.}

\subsection{Case Study: Migrating to Different Backends}\label{sec:res-backend}

\textbf{Target and metric.} This case study again uses LoRA+~\cite{hayou2024loraplus}, but reproduces the paper's GPT-2 experiment on MNLI, where LoRA+ outperforms the best tuned LoRA in validation accuracy on MNLI-matched (82.0\% vs.\ 81.3\%, \goodnum{$+0.7$ percentage points}). As before, we run only LoRA+ and use the paper's value for LoRA. The original code reaches 81.39\% on an NVIDIA L4 GPU, and we use this run as the reference: a migration preserves the claim (L3) if its TPU result exceeds the GPU's.

\textbf{Design.} A hardware migration fixes the target chip but not the software stack that drives it. We compare PyTorch/XLA and TorchTPU on the same single-chip TPU v5e-1, preserving the global batch size (32), training steps (36,816), and other hyperparameters. PyTorch/XLA compiles lazy-tensor graphs, whereas TorchTPU, open-sourced by Google and Meta in October 2026~\cite{basile2026torchtpu}, executes operations eagerly by default on a native \texttt{tpu} device. The environments also use different PyTorch versions (2.6 and 2.14), so the comparison does not isolate backend effects.

\textbf{Results (L1, L3).} Both backends stay close to the GPU in accuracy (Fig.~\ref{fig:pilot-targets}a), and the training-loss curves of all three runs nearly coincide (Fig.~\ref{fig:pilot-targets}b). PyTorch/XLA ends at 81.80\% on MNLI-matched, above the GPU, and passes L3. TorchTPU reaches 82.70\% on MNLI-mismatched, slightly above the GPU's 82.66\%, but 81.23\% on MNLI-matched, 0.16 percentage points below the GPU and just below the paper's 81.3\% for plain LoRA, so by our criterion L3 fails. However, this small difference does not establish a backend-induced defect: TorchTPU's final training loss closely matches the GPU's (0.458 versus 0.456 over the last 500 steps), and its MNLI-mismatched accuracy is slightly higher. The two TPU backends differ by an average of 0.50\% across 73 evaluations, but single-seed executions and different PyTorch versions prevent attributing this difference to the backend alone. For practitioners, such borderline results illustrate why acceptance thresholds must account for workload requirements and numerical variability.

\textbf{Migration on each backend.} TorchTPU required fewer changes in this case: the agent added three files ($+$92 lines), with only a 28-line launcher containing executable code, and modified no original files. Its main compatibility problems involved pinned \texttt{transformers} and \texttt{accelerate} versions that predated the native \texttt{tpu} device and silently placed the model and batches on the CPU when CUDA was unavailable. The agent identified and fixed these problems. The PyTorch/XLA migration changes five files ($+$135/$-$1 lines), including one original file, with 68 lines of code: a launcher that starts one process per chip and a trainer that fixes a checkpoint problem caused by lazy tensors. These differences suggest that backend interfaces can affect migration effort, but also show that apparently simpler device support does not eliminate the need to verify where computation actually executes.

\textbf{Cost and performance (L4).} TorchTPU is the fastest and cheapest of the three (Fig.~\ref{fig:pilot-targets}c--d). Training takes 1.98 hours, 1.86 times faster than the GPU (3.68 hours) and 1.23 times faster than PyTorch/XLA (2.44 hours), and one run costs \$2.37, 25\% less than on the GPU (\$3.14) despite the TPU's 41\% higher hourly price. TorchTPU also keeps the chip busy 88\% of the training time, against 11\% under PyTorch/XLA, where the chip mostly waits for the host to trace and dispatch the graph (Fig.~\ref{fig:pilot-targets}e). However, the eightfold difference in TPU duty cycle translates into only a 1.23-fold speedup. This indicates that utilization alone does not determine throughput: the backends differ in how efficiently they execute work while the accelerator is active. For engineering teams, both execution efficiency and hourly pricing matter when comparing the operational costs of migrated workloads.

\textbf{Room for optimization.} The agent migrated but did not optimize, leaving backend-specific bottlenecks: PyTorch/XLA spends about 186~ms per step tracing and dispatching against 26~ms of chip work, and fusing GPT-2's GELU cuts TorchTPU's step from 150 to 112~ms (25\%), while PyTorch/XLA barely changes.

\findingbox{Finding.}{On the same TPU hardware, TorchTPU required fewer code changes and completed training 1.86 times faster and 25\% cheaper per run than the GPU, although its single-seed accuracy narrowly missed the study's L3 criterion. A targeted operator fusion further reduced fixed-batch step time by 25\%. Backend selection therefore requires separate verification of model behavior, migration effort, and execution cost; higher performance alone does not establish an acceptable migration.}

\subsection{Implications for Practice}\label{sec:implications}

Although we study open-source ML research projects rather than production systems, their techniques, including parameter-efficient fine-tuning and model compression, are also relevant to industrial ML workloads. These repositories expose practical migration challenges: only seven of the 148 collected projects contained author-written tests, and dependency compatibility repeatedly complicated migration. Our findings therefore inform engineering practice, although they do not establish how frequently these problems occur in production. We discuss implications for practitioners adopting coding agents and tool builders developing them.

\textbf{For practitioners.}
\emph{Define acceptance criteria and establish an independent reference before migration.} Successful execution and agent-reported success do not guarantee preserved behavior (Section~\ref{sec:res-hardware}). Teams should determine which consistency dimensions matter for their workload (L0--L4) and define acceptable numerical and performance differences. They should also retain a GPU reference under comparable settings and seeds, checking training curves and evaluation metrics rather than only final results. When the original execution already misses its expected outcome, teams must distinguish inherited discrepancies from migration-induced changes. For stochastic tasks, deterministic comparisons of output distributions can be more informative than fixed thresholds on sampled metrics, as observed in LM-Steer.

\emph{Verify execution semantics under the intended hardware configuration.} Migration failures may arise from unchanged dependencies rather than agent-modified code (Section~\ref{sec:rq2}). Beyond confirming device placement and the absence of CPU fallback or unexpected recompilation, teams should inspect checkpoint restoration, optimizer state, effective batch size, and gradient operations. The four-chip LoRA+ case demonstrates why these checks must be repeated when changing chip counts or backends: a migration validated on one chip can silently alter distributed optimization on multiple chips.

\emph{Evaluate total migration cost, not only agent tokens or execution speed.} Migration does not guarantee faster or cheaper training. TorchTPU reduced one workload's training time by 1.86$\times$ and per-run cost by 25\%, whereas TPU/GPU run costs varied from 0.47 to 3.6 times across projects. Agent token charges (US\$2--US\$23 per project) are only one component; failed attempts, verification runs, and human supervision (2.4 interventions per migration on average) also matter. Teams should therefore evaluate expected operational savings against migration and validation effort rather than infer economic benefits from a successful run.

\textbf{For migration tool builders.}
\emph{Integrate evidence-based verification into the migration workflow.} Tools should verify execution integrity, numerical agreement, and project-specific goals before reporting success, rather than relying on agent self-assessment: the agent stopped excessively slow attempts but did not identify silent numerical or behavioral drift (Section~\ref{sec:rq3}), and in the four-chip case it reported success despite a silent accuracy regression (Section~\ref{sec:res-hardware}). Each conclusion should be supported by GPU/TPU comparisons and runtime evidence: device placement, CPU fallbacks, compilation activity, training curves, projected runtime, and operations whose semantics can change across devices, such as checkpoint restoration and gradient reduction before clipping.

\emph{Encode accelerator- and dependency-specific failure knowledge.} Repeated recompilation, host synchronization, stale parameter references, and distributed gradient-clipping behavior require knowledge beyond ordinary code translation. Migration tools could incorporate checks and known fixes for these patterns, including avoiding data-dependent shapes, reducing unnecessary host synchronization, rebinding optimizer parameters when needed, and verifying gradient reduction order. The agent's incorrect diagnosis of the four-chip regression shows why this knowledge should guide diagnosis rather than defaulting to familiar explanations such as numerical precision.

\emph{Make stopping criteria configurable and evidence-driven.} Smoke tests catch problems before expensive full runs, but fixed thresholds produced six false alarms among seven smoke-test stops (Section~\ref{sec:rq3}), triggered by loss improvements, inconsistent reference setups, and stochastic outputs. Tools should let users specify acceptance thresholds, computational budgets, and when human approval is required; distinguish improvements from regressions; account for stochastic evaluation; exclude idle or approval-waiting time from budgets; and, when the evidence is insufficient, explain the uncertainty and request a decision instead of declaring failure.

\emph{Separate migration correctness from performance tuning.} Seven of the nine completed TPU runs had accelerator utilization below 25\%, while backend selection and operator fusion revealed further optimization opportunities. In the TorchTPU case, fusing GELU reduced fixed-batch step time by 25\% (Section~\ref{sec:res-backend}). A tool could therefore treat optimization as a subsequent stage, identifying host overhead or unfused operators after consistency checks pass, and rerunning those checks after each optimization to avoid introducing new behavioral changes.

\emph{Maintain reviewable migration traces.} Each attempt's code changes, expected and claimed levels, verified outcomes, failure diagnoses, and human decisions provide evidence for acceptance and material for improving migration tools. Such records can support diagnosis benchmarks, better stopping rules, and upstream dependency fixes. Our replication package provides these traces~\cite{sun2026seipreplication}.

\section{Threats to Validity}\label{sec:threats}

\textbf{Construct validity.} Our levels, and the checks and thresholds behind them, are one way to define consistency. Our L3 judgment relaxes the requirement and could let a weaker migration pass, and another analyst might classify some failures differently. Each run uses a single seed, so differences below 1\% between the code before and after migration cannot be distinguished from run-to-run variation; we therefore report L1 as a measurement rather than a verdict. To make these judgments checkable, the replication package contains the evidence for every judgment, including the full trace of every attempt. Costs use on-demand list prices, and the share of time the accelerator is busy comes from sampled measurements; L4 therefore depends on these prices, the GPU model, and a single run, and part of a speedup may come from code changes the migration required rather than from the hardware.

\textbf{Internal validity.} The agent is nondeterministic, and a second session could take a different path for the same project. The agent reads the level definitions before migrating, which may make it check its own work more carefully than an agent without them would; its claims, however, are judged against our measurements, not taken at face value. The experimenter's decisions, such as approving a full run after a smoke-test gate, influence the outcomes; we count every intervention and report it per project. Migrating also changes the software stack: the agent upgrades pinned libraries to versions that support the target, so part of a difference may stem from a dependency version rather than the migration itself.

\textbf{External validity.} We study one agent (Claude Code with Claude Opus~5.5), one target (a single TPU v5e chip, mainly with PyTorch/XLA), PyTorch projects only, and ten research projects that fit on one chip; production code, larger models, multi-chip targets, and other agents or accelerators may behave differently. The candidates were collected with LLM-assisted search, which may miss projects that meet our criteria, and we preferred less widely adopted projects to reduce training-data contamination, so our sample may be harder or easier to migrate than popular projects. The software stack evolves quickly: TorchTPU was released after our study design was fixed, and newer versions of the backends and libraries may remove some of the failures we observed. With ten projects, our results are descriptive; we do not claim statistical generalization, and we release the protocol so that others can extend the study.

\section{Conclusion and Future Work}\label{sec:future}

We used Claude Code to migrate ten PyTorch research projects from GPUs to TPUs and verified each migration level by level with the L0--L4 consistency hierarchy. After migration, 8 projects reach L3, 3 of them by matching GPU runs that themselves miss the paper's claim, with 2.4 human interventions per migration on average. Yet code that runs is not a successful migration: the agent was often too optimistic before an attempt, 49\% of the failures raised no error, and the hardest problems to find came from the interaction between the migrated code and third-party libraries. The agent itself is inexpensive, but whether a migration pays off depends on the target and its configuration. Agent-driven hardware migration therefore needs independent, execution-based verification.

\textbf{Future work.} We plan to (1) turn the migration and verification process into a tool that migrates the replication packages of newly published papers and submits only migrations that reach L3 to the authors as pull requests; (2) design early-stopping rules based on the required level, the target hardware, and the budget; (3) use the recorded migration traces to build a benchmark for failure diagnosis and a knowledge base of migration fixes; and (4) extend the study into a SWE-bench-style benchmark~\cite{jimenez2024swe} for cross-hardware ML migration, in which each task asks an agent to migrate a research project to a target accelerator, and the consistency hierarchy scores the result level by level against reference runs on the original hardware.

\clearpage
\bibliographystyle{IEEEtran}
\bibliography{references}

\end{document}